\documentclass[a4paper, 12pt]{paper}

\usepackage{microtype}
\usepackage[Lenny]{fncychap}
\usepackage{tikz}
\usepackage{pgfplots}
\usepackage{amssymb, amsthm, amsmath}

\usepackage{latexsym,multicol,graphicx}
\usepackage{euscript}
\usepackage{ stmaryrd }
\usepackage{fancyhdr}
\usepackage{accents}
\usepackage{graphicx, verbatim}
\usepackage{xcolor}
\usepackage{verbatim}

\usepackage[english]{babel}

\usepackage[a4paper, hmargin=3cm, vmargin={4.5cm, 4.5cm}]{geometry}
\usepackage{enumerate}
\usepackage{fancyhdr}
\usepackage{relsize}

\newcommand{\diagdots}[3][-25]{%
  \rotatebox{#1}{\makebox[0pt]{\makebox[#2]{\xleaders\hbox{$\cdot$\hskip#3}\hfill\kern0pt}}}%
}

\usepackage[toc,page]{appendix}

\newtheorem{teo}{Theorem}[section]
\newtheorem*{teo*}{Theorem}

\newtheorem{obs}{Observation}[section]

\newtheorem{cl}{Claim}
\theoremstyle{definition}

\newtheorem*{rmk}{Remark}

\renewcommand{\phi}{\varphi}

\makeatletter
\newcommand\opn{\mathrel{\opncls@{\circ}}}
\newcommand\cls{\mathrel{\opncls@{\bullet}}}
\newcommand{\opncls@}[1]{%
  \vphantom{\subseteq}
  \mathpalette\opncls@@{#1}%
}
\newcommand{\opncls@@}[2]{%
  \ooalign{$\m@th#1\subseteq$\cr
  \hidewidth\opncls@fix{#1}\hbox{$\m@th#1#2\mkern.5mu$}\cr}}

\newcommand\opncls@fix[1]{%
  \ifx#1\displaystyle
    \raise.165ex
  \else
    \ifx#1\textstyle
      \raise.165ex
    \else
      \ifx#1\scriptstyle
        \raise.115ex
      \else
        \raise.085ex
      \fi
    \fi
  \fi
}
\makeatother
\DeclareMathOperator{\arccosh}{arccosh}

\newcommand{\diam}{\mathrm{diam}}
\newcommand{\Res}{\mathrm{Res}}

\newcommand{\e}{\mathrm{e}}

\newcommand{\one}{\mathbf{1}}

\newcommand{\R}{\mathbb R}
\newcommand{\E}{\mathbb E}

\newcommand{\Z}{\mathbb Z}

\newcommand{\PP}{\mathbb P}

\newcommand{\kk}{\kern 0.1em}

\usepackage[notext, upint]{stix}
\usepackage{step}
\usepackage[T1]{fontenc}

\usepackage[colorlinks]{hyperref}

\begin{document}
\title{Non-Lyapunov annealed decay for 1d Anderson eigenfunctions}
\author{Davide Macera\textsuperscript{1} \textsuperscript 2}
\maketitle
\begin{abstract} 
In \cite{JKL} Jitomirskaya, Kr\"uger and Liu analysed the dynamical decay in expectation for the supercritical almost-Mathieu operator in function of the coupling parameter , showing that it is equal to the Lyapunov exponent of its transfer matrix cocycle, and asked whether  the same is true for the 1d  Anderson model. We show that this is essentially never true when the disorder parameter is sufficiently large.
\end{abstract}
\footnotetext[1]{Department of Mathematics, Durham University, Durham, UK.  Email: davide.macera@durham.ac.uk.\\  Supported by EPSRC EP\textbackslash T004290\textbackslash 1}
\footnotetext[2]{This work was started when the author was a PhD student at Roma Tre university. }
\section{Introduction}
Consider the one-dimensional Anderson model, i.e.\ the operator $H$ acting on a dense subset of $\ell^2(\mathbb Z)$ via
\begin{equation}\label{eq:defop}
(H \psi)(x) =\psi(x+1) + V_x \psi(x) +  \psi(x-1)~, \quad x \in \mathbb Z~,
\end{equation}
where $V_x$ are i.i.d. random variables. We assume that the distribution of $V_0$ is bounded and not concentrated at one point (in most of the discussion below, the first assumption can be relaxed to the existence of a finite fractional moment $\mathbb E |V_0|^\eta < \infty$). Carmona--Klein--Martinelli \cite{CKM} showed that under these assumptions $H$ exhibits Anderson localisation, i.e.\ almost surely $H$ has pure point spectrum, and moreover 
\begin{equation}\label{eq:ef-dec} \mathbb P \left\{ \forall (\lambda, \psi) \in E \,\,\, \limsup_{x \to \pm \infty} \frac{1}{|x|} \log |\psi(x)| = - \gamma(\lambda) \right\} = 1~,\end{equation}
where $\mathcal E[H]$ is the collection of eigenpairs of $H$ (the spectrum is almost surely simple, so $|\psi|$ is well-defined), and $\gamma(\lambda)$ is the Lyapunov exponent of $H$ at energy $\lambda$.  Under more restrictive assumptions on $H$, the pure point nature of the spectrum was first proved by Goldsheid--Molchanov--Pastur \cite{GMP} and by Kunz-Souillard \cite{KS}; the exponential decay of the eigenfunctions was first established by Molchanov \cite{Molch}.  

While the proof of \cite{CKM} employs multi-scale analysis, single-scale proofs have recently been found by Bucaj et al.\ \cite{BDFGVWZ}, Gorodetski--Kleptsyn \cite{GorKl}, and Jitomirskaya-Zhu \cite{JZ}. Generalisations to models with off-diagonal disorder and to matrix-valued potentials are studied in \cite{R,MS}. 

A stronger notion of Anderson localisation involves the notion of eigenfunction correlator, introduced by Aizenman \cite{A}. Denote
\[ Q(x,y) = \sup \left\{ |f(H)(x,y)| \,\, : \,\, f : \mathbb R \to \mathbb C~, \,\, \sup |f| \leqslant 1 \right\}~,\] 
where the supremum is taken over Borel functions. If $H$ has pure point spectrum, the correlator takes the form
\[ Q (x, y) =  \sum_{(\lambda, \, \psi) \kk \in \kk E} |\psi(x)| |\psi(y)|~.\]
Then there exists $\gamma > 0$ such that for any $x$
\begin{equation}\label{eq:q-dec} \mathbb P \left\{  \limsup_{y \to \pm \infty} \frac{1}{|y-x|} \log Q(x, y) \leqslant -\gamma \right\} = 1~.\end{equation}
In fact, in the current setting (\ref{eq:q-dec}) holds with $\gamma = \gamma_{\inf}$, where
\begin{equation}\label{eq:inf}
\gamma_{\inf} = \inf_{\lambda \kk \in \kk \sigma(H)} \gamma(\lambda)
\end{equation}
and $\sigma(H)$ is the spectrum of $H$ (a deterministic set) -- see \cite{JZ}. This strong form of (\ref{eq:q-dec}) implies (\ref{eq:ef-dec}), as well as dynamical localisation, decay of the Fermi projection as well as other properties of relevance in quantum dynamics.

Ge and Zhao built on the work \cite{JZ} and proved the following :
\begin{teo}[Ge--Zhao]\label{thm:gz}
For the operator $H$ of (\ref{eq:defop}) with $V_0$  bounded and not concentrated at one point, one  has, for any $x \in \mathbb Z$,
\begin{equation}\label{eq:q-dec-mean}  \gamma^{\mathbb E} =- \limsup_{y \to \pm \infty} \frac{1}{|y-x|} \log \mathbb E Q(x, y)  > 0~.\end{equation}
\end{teo}
In Section~\ref{s:gz} we give another, arguably, simpler, proof of this result, adopting an argument from \cite{ESS}.

\medskip
Jitomirskaya, Kr\"uger and Liu \cite{JKL} studied the validity of (\ref{eq:q-dec-mean}) in the almost-periodic setting, namely, for the supercritical almost-Mathieu operator with Diophantine frequency, and showed that in that setting $\gamma^{\mathbb E}$ can be taken to be equal to $\gamma_{\inf}$. They asked whether the same is true for the Anderson model. We show that this is not the case. A first counterexample comes from the Anderson-Bernoulli model:
\begin{teo}\label{thm:2} For $a > 0$, consider the operator $H_a=H_0+ aV$ with $V_x$ being a random variable having an atom at $0$.. Then  $\gamma^{\mathbb E}$ is bounded from above uniformly in $a$. 
\end{teo}
In particular, if $V_x$ is a Bernoulli random variable with parameter $p$,  by a result of Martinelli and Micheli \cite{MM}, $\gamma_{\inf} \geqslant c \, \log a$ for sufficiently large $a$. Therefore, by the above theorem, $\gamma^\E(H^a) \neq \gamma_I(H^a)$ for $a$ large enough.\\
Furthermore, the above theorem remains true for \emph{any} bounded random potential with finite first moment,  at sufficiently high disorder:
\begin{teo} \label{teo:3}
Let $V=\{V_i\}_{i \in \Z}$ be a nondeterministic,  bounded, i.i.d. random potential  such that $\E[V_0] < \infty$ , and let $H^a:= H_0+a V$. \\ Then, for any $a$ large enough
\begin{equation*}
\inf_{E \in \kk \sigma(H^a)} \gamma_E(H^a) > \gamma^\E(H^a) ~.
\end{equation*}
\end{teo}
\paragraph{Acknowledgements} 
I'm deeply indebted to my former PhD advisor, Sasha Sodin, for suggesting the topic of this paper and for giving major contributions to its development.  I believe he should have been a co-author of this paper, but for reasons I don't understand he asked me to erase his name from it.\\
 I'm also grateful to Alexander Elgart for many useful comments on a preliminary version of this paper.

\section{Proof of Theorem~\ref{thm:gz}}\label{s:gz}
For $a, b \in \mathbb Z$, denote by $H_{[a,b]}$ the restriction of $H$ to $[a, b]$ (with Dirichlet boundary conditions), and let $G_E[H_{[a,b]}] = (H_{[a,b]}-E)^{-1}$.  Let $\tau > 0$, $E \in \mathbb R$ and $N \geq 1$. A site $x \in \mathbb Z$ is called $(\tau, E, N)$-nonresonant ($x \notin \Res(\tau, E, N)$) if 
\[ |G_E[H_{[x-N,x+N]}](x, x\pm N)|\leqslant \e^{-(\gamma(E) - \tau)N}~. \]
Otherwise, $x$ is called $(\tau, E, N)$-resonant ($x \in \Res(\tau, E, N)$). The proof of the theorem uses the following 
\begin{cl}\label{cl:1} Assume that $V_0$  is bounded and not concentrated at one point. Then for any $\tau > 0$ there exist $C, c> 0$ such that 
\[ \mathbb P \left\{ \forall E \in \mathbb R \,\, \diam \Res(\tau. E,N) \cap [-N^2, N^2] > 2N\right\} \leqslant C\e^{-cN}~.\]
\end{cl}
See \cite[Proposition 2.1]{MS} for this formulation (in the more general case of matrix potentials) and \cite{JZ} for similar statements.

Next, we need a representation for the eigenfunction correlator as a singular integral (see \cite{AW}):
\begin{equation}\label{eq:qrepr}
Q(x,y) = \lim_{L \to \infty} Q^L(x,y)~, \quad Q^L(x,y) = \lim_{ \epsilon \to 0^+} \frac{\epsilon}{2} \int |G_E[H_{[-L,L]}](x, y)|^{1-\epsilon}\mathrm{d}E \leqslant 1~.
\end{equation}

Having these two ingredients, we argue as follows. Without loss of generality we can assume that $x = 0$. Set $\tau = \frac12 \min_{E \in \sigma(H)} \gamma(E)$ and $N = \lfloor \frac y {10} \rfloor$, and consider the event
\[ \mathcal R = \left\{ \exists E \in \mathbb R \,\, : \,\, 0, y \in \Res(\tau, E, N) \right\}~. \]
According to Claim~\ref{cl:1}~, $\mathbb P(\mathcal R) \leqslant C \e^{-cN}$. 
On the complement $\Omega \setminus \mathcal R$, we have $\mathbb R = A \cup B$, where
\[ A = \left\{ E \in \mathbb R \,\,  : \,\, 0 \notin \Res(\tau,E,N)\right\} ~, 
\quad B = \left\{ E \in \mathbb R \,\,  : \,\, y \notin \Res(\tau,E,N)\right\}~.\]
For $E \in A$, 
\begin{equation*}
\begin{split}
G_E[H_{[-L,L]}](0, y) =  \, & G_E[H_{[-N,N]}](0, N)  G_E[H_{[-L,L]}](N+1, y) \, +\\
 +  \, & G_E[H_{[-N,N]}](0, -N)  G_E[H_{[-L,L]}](-N-1, y)~, 
\end{split}
\end{equation*}
whence
\[ |G_E[H_{[-L,L]}](0, y)| \leqslant \ \e^{-\tau N} (| G_E[H_{[-L,L]}](N+1, y) | + |G_E[H_{[-L,L]}](-N-1, y)|~. \]
and an analogous bound can be deduced for $B$.\\
Thus
\[ \lim_{\epsilon \to +0} \frac{\epsilon}{2} \bigg[\int_{A} + \int_B \bigg] |G_E[H_{[-L,L]}](0, y)|^{1-\epsilon} \mathrm{d}E \leqslant 4 \e^{-\tau N}~.\]

Finally,
\[\begin{split} \mathbb E Q(0,y) 
&= \mathbb E (Q(0, y) \kk | \kk \mathcal R) \kk  \mathbb P(\mathcal R) +  \mathbb E (Q(0, y) \kk | \kk \Omega \setminus \mathcal R) (1 - \mathbb P(\mathcal R))
\\
&\leqslant \mathbb P(\mathcal R) + \mathbb E (Q(0, y) \kk  | \kk \Omega \setminus \mathcal R)  
\leqslant C \e^{-cN} + 2\e^{-\tau N}~. \end{split}\]
Thus $\gamma^{\mathbb E} \geqslant \min(c,\tau) >0$.
\begin{rmk} This proof can be extended  to  quasi-one-dimensional operator, such as the Anderson model on the strip of width $W$ or the more general model studied in \cite{MS}. A slightly weaker version of (\ref{eq:qrepr}) is still true in this case (see \cite{ESS}):
\begin{equation*}
Q^L(x,y) \leqslant  \lim_{ \epsilon \to 0^+} \frac{\epsilon}{2} \int \|G_E[H_{[-L,L]}](x, y)\|^{1-\epsilon}\mathrm{d}E \leqslant W~.
\end{equation*}
and the argument above follows with minor modifications.
\end{rmk}
\section{Proof of Theorem~\ref{thm:2}}

Let $K > 0$ be a large numerical constant (independent of any parameters), to be specified later. For $x > 0$, consider the event 
\[ \Omega_{K,x}= \left\{ \forall x \in [-K x,  (K+1)x ], \,\, V_x  = 0\right\}~.\]
 We shall prove the following: for any $\epsilon > 0$, one has on $\Omega_{K,x}$ for sufficiently large $x$:
\begin{equation}\label{eq:need2}
 Q(0, x) \geqslant \e^{-\epsilon | x|}~.
\end{equation}
Since $\PP(\Omega_{K,x}) =(1-p)^{2Kx+1}$, this this would imply that
\begin{equation}\label{eq:limp}
\gamma^\E \leqslant -\lim_{x \to \infty}  \dfrac{1}{x} \log(1- p)^{2Kx+1} \e^{-c x} \leqslant c- 2K \log(1-p)
\end{equation}
as claimed.

We now turn to the proof of (\ref{eq:need2}). Observe that for any $\delta > 0$
\begin{equation} \label{eq:im}
 Q(0,x) \geqslant \delta \, |G_{\delta}[H](0, x)| \geqslant  \delta|G_{\delta}[H-2 \cdot \one](0, x)|~. 
\end{equation}
Let $T = H_0$ be the free Laplacian (obtained by setting $V_x \equiv 0$ in (\ref{eq:defop})), and let $T_{K,x}$ be the restriction of $T$ to the finite volume $[-Kx, (K+1)x]$. Then 
\[\begin{split} &|G_{\delta}[H- 2 \cdot \one](0, x)| \geqslant \\
&\quad \geqslant |G_{\delta}[T_{K,x}-2 \cdot \one](0, x)| - \\
&\quad\qquad-  |G_{\delta}[T_{K,x}-2 \cdot \one ](0, -Kx)| |G_{\delta}[H-2 \cdot \one]( -Kx-1, x )| - \\
&\quad\qquad-  |G_{\delta}[T_{K,x}-2 \cdot \one](0,  (K+1)x)| |G_{\delta}[H-2 \cdot \one]( (K+1)x+1, x )| \\
&\quad \geqslant |G_{\delta}[T](0, x)|- \\
&\quad\qquad-  |G_{\delta}[T_{K,x}-2 \cdot \one](0, -Kx)| |G_{\delta}[H-2 \cdot \one]( -Kx-1, x )|- \\
&\quad\qquad-  |G_{\delta}[T_{K,x}-2 \cdot \one](0,  (K+1)x)| |G_{\delta}[H-2 \cdot \one]( (K+1)x+1, x )| -\\
&\quad\qquad-  |G_{\delta}[T_{K,x}-2 \cdot \one](0, -Kx)| |G_{\delta}[T-2 \cdot \one]( -Kx-1, x )|- \\
&\quad\qquad-  |G_{\delta}[T_{K,x}-2 \cdot \one](0,  (K+1)x)| |G_{\delta}[T-2 \cdot \one]( (K+1)x+1, x )|~.
\end{split}
\]
By the Combes--Thomas estimate (\cite{AW}, Theorem 10.5, statement 2), noticing that $H-2\cdot \one$ and $T-2\cdot \one$ are negative operators, and thus $\delta$ sits above their spectra, we deduce that for $\delta \in (0, 1)$ 
\[ |G_{\delta}[H-2 \cdot \one](0, x)| \geqslant |G_{\delta}[T-2 \cdot \one](0, x)| - \tfrac{C_1}{\delta} \kk \e^{-c_1(2K+1)\sqrt{\delta}\kk | x |}~. \]
Now, 
\begin{equation}\label{eq:lb}
|G_{\delta}[T-2 \cdot \one](0, x)|  \geqslant C_2 \e^{-c_2 \sqrt{\delta} \kk| x|}~.
\end{equation}
Indeed,  $g(x) = G_{\delta}[T+2 \cdot \one](0,x)$ is by definition the square-summable solution to the equation
\[ g(x+1) + g(x-1) +(2+\delta) g(x) = \delta_{x,0}~.\]
Plugging in the ansatz $g(x) =\alpha \e^{-\xi|x|}$, we find that this is indeed a solution provided that 

\[ \xi= \arccosh \Big(1+\dfrac{\delta}{2}\Big); \qquad \alpha = (2+2\e^{-\xi}+\delta)^{-1}\]
hence $|g(x)| \geqslant |\alpha|\e^{ \kk -\xi |x|}$, where $\xi \geqslant c \sqrt{\delta}$. Having 
 set $\delta = (\epsilon/c_2)^2$ and $2K+1 =\lceil 100 \kk c_2/c_1 \rceil$, we obtain (\ref{eq:need2}).
\qed
\section{Proof of Theorem \ref{teo:3}: logarithmic divergence for general potentials}
\label{sec:CS}
A version of the Martinelli-Micheli bound the 1d Anderson model with absolutely continuous, bounded potential has been proven in 1983 by Avron, Craig and Simon in \cite{ACS}.
\begin{teo}\textup{(Avron, Craig, Simon, \cite{ACS})}\\
Let $H^a=H_0+a V$ be a random Schr\"odinger operator where $V$ is a bounded random  potential with absolutely continuous density. Then the Lyapunov exponent $\gamma_a$ of $H_a$ is such that
\begin{equation*}
\gamma_a \geqslant \log(a)-K
\end{equation*}
where $K$ is a finite constant. 
\end{teo}
We will adapt the proof in \cite{ACS} to  any potential having finite first moment (not necessarily absolutely continuous). Avron, Craig and Simon's proof relies on the  Thouless' formula for the Lyapunov exponent of a Schr\"odinger operator $H$,  stating that
\begin{equation} \label{eq:thou}
\gamma_H(E)= \int \log|E-E'|\, \mathrm{d} \kappa_H(E')
\end{equation}
where $\kappa_H(E')$ denotes the \emph{integrated density of states} of the operator $H$. They proceed then to bound the negative part of the logarithm in \ref{eq:thou} using the \emph{Wegner estimate}: If $H=H_0+V$ is a random Schr\"odinger operator with i.i.d. potential, then
\begin{equation*}
\dfrac{\mathrm{d}\kappa_H(E)}{\mathrm{d}E} \leqslant \|\rho_V\|_{\infty}
\end{equation*}
which is unfortunately proven true  only when the  distribution $\rho_V$ of the potential is absolutely continuous.\\ Fortunately, Shubin, Vakilian and Wolff proved in \cite{SVW} a slightly weaker bound for the IDS of a random Schr\"odinger operator whose potential satisfies the conditions of Theorem \ref{teo:3}: if $\E[|V_0|]  < \infty$, then
\begin{equation} \label{teo:SVW}
|\kappa_H(E)-\kappa_H(E')| \leqslant C|E-E'|^\alpha
\end{equation}
for some $\alpha \in (0,1)$ and some constant $C>0$. This bound is sufficient to let the Avron-Craig-Simon argument work in the present generality. \\
By applying the Thouless formula  to $H^a$ and splitting the logarithm into its positive and negative parts, we get that
\begin{equation*}
\begin{split}
 &\int \log|E-E'|\, \mathrm{d} \kappa_{H^a}(E')  =\\ 
= \ & \int \log_+|E-E'|\, \mathrm{d} \kappa_{H^a}(E')-\int \log_-|E-E'|\, \mathrm{d} \kappa_{H^a}(E')  \\
\geqslant  \ & \int_{\kk a/2}^\infty \log_+|E-E'|\, \mathrm{d} \kappa_{H^a}(E')-\int_{ \kk 0}^\infty \kk \kappa_{H^a} \{E' :  |E-E'| \leqslant \e^{-t}\} \kk  \mathrm{d} t
\end{split}
\end{equation*} 
We claim that
\begin{equation} \label{eq:ub}
\int_{ \kk 0}^\infty \kk \kappa_{H^a} \{E' :  |E-E'| \leqslant \e^{-t}\} \kk  \mathrm{d} t \leqslant  C ~;
\end{equation}
for some $C>0$ uniform in $a$, and that 
\begin{equation} \label{eq:lb}
\int_{\kk a/2}^\infty \log_+|E-E'|\, \mathrm{d} \kappa_{H^a}(E') \geqslant c \log a ~.
\end{equation}
for some positive constant $c$.\\
The first bound (\ref{eq:ub}) is  proven by using  inequality  \ref{teo:SVW} :
\begin{equation*}
\begin{split}
\int_{ \kk 0}^\infty \kk \kappa_{H^a} \{E' :  |E-E'| \leqslant \e^{-t}\} \kk  \mathrm{d} t \leqslant \ & \|V\|_{\infty}\int_{\kk 0}^\infty\ \e^{-t} \mathrm{d} t^\alpha \\
\leqslant \ &  C \int_0^\infty \alpha t^{\alpha -1} \e^{-t} \mathrm{d} t  \\
\leqslant \ &C\kk \Gamma(\alpha)\leqslant C'~.
\end{split}
\end{equation*}

\begin{rmk} We strongly believe that the main result of this section  (and thus Theorem \ref{teo:3} as a whole) can be extended  to the most general setting for which 1-d localisation has been proven (nondeterministic potential with any finite fractional moment). However, generalising [\ref{teo:SVW}] to the case where one only has a generic fractional moment  appears to be nasty . \\ An alternative to this approach could be extending the proof of the logarithmic divergence  for the Anderson-Bernoulli model in \cite{MM} to the general case; however, even if  this seems to be doable and should not present major technical difficulties, additional estimates would be needed to make Martinelli's and Micheli's already six page long proof work for generic potentials, and many formulas would get much longer and nastier. \\
In conclusion, the length of the present paper would likely get doubled by such an attempt, therefore we avoid it to keep the paper short and more readable while keeping the result reasonably general.
\end{rmk}
\section{Proof of Theorem \ref{teo:3}:  Green function estimates}

Since the above argument uses crucially the fact that a Bernoulli random variable is $0$ with positive probability, one might suspect that the presence of an atom at zero is required for the annealed dynamical decay to be non-Lyapunov. However, in this section we will use a simple trick to generalise the result to any bounded potential.The trick relies on the observation that it is possible to decompose any dilated    random variable  $aX$ having an atom as a sum of   two (not necessarily independent ) random variables, one of which is bounded in $a$, and the other has basically the same distribution as $aX$ with the difference that the atom has been subtracted some mass. If we subtract in this way mass from the atom at a sufficient rate, and control the error given by the bounded addend, then we can show that the growth of the annealed decay rate in $a$ is much slower than the logarithmic lower bound prescribed by the the results of Avron-Craig-Simon and Shubin-Vakilian-Wolff. We first state a relevant result by Avron, Craig and Simon on logarithmic divergence in the coupling of the Lyapunov exponent for 1d Anderson models that have absolutely continuous bounded potentials.

\begin{proof} 

 In order to exploit  the Bernoulli case, we will make use of the following observation, 
 
 \begin{obs} Let $X$ be a bounded random variable, and let $\bar{x}\in \mathrm{supp}(X)$. Suppose that $X$ is absolutely continuous in a neighborhood of $\bar{x}$ and denote by $\widetilde{X}^\epsilon$ the random variable having the same density as $X$, except for the fact that it has an atom at $\bar{x} $ of mass $\epsilon$, suitably renormalised. Then there exists a bounded random variable $\tilde{\eta}$ (not necessarily independent on $X$) such that
 \begin{equation*}
 a X\stackrel{\mathrm{d}}{=}a'\widetilde{X}^\epsilon +\tilde{\eta}
 \end{equation*}
 for some $a' >0$.
 \end{obs}
 This observation basically asserts that we can remove (or, by extension, subtract mass to) an atom from the distribution of a random variable at the cost of adding another (dependent) random variable uniformly bounded  in the coupling. \\
 It follows by simply observing that if $\tilde\eta$ has the same distribution of $X$ and $\widetilde{X}^\epsilon $ is chosen to take the same values as $\tilde \eta$ (so that $\widetilde{X}^\epsilon$ would retain its usual law and its atom at $\bar{x}$, but becoming totally dependent on $\tilde\eta$), then $a\widetilde{X}^\epsilon+\tilde\eta  \stackrel{\mathrm{d}}{=} (a+1) X$. \\
We use this observation on the potential $V$ with $\epsilon=\epsilon(a) \gg a^{-\beta}$ for all $\beta >0$, and decompose $aV=a\widetilde{V}^{\epsilon(a)}+ \tilde{\eta}_R$, where $\tilde{\eta}_R$ is supported on the interval $(-R, R)$. Then  $H_a=T^{R}+a\widetilde{V}^{\epsilon(a)}$ where $T^{R}=T_0+\tilde\eta_R$ , and $\widetilde{V}^{\epsilon(a)}$ is a random variable having the same distribution as $V\stackrel{\mathrm{d}}{=} \tilde{\eta}_R$ except for having an atom at $0$ of mass $\epsilon(a)$. \\
Then again, if $Q(0, x) \geqslant \e^{-\epsilon  | x|}$ , we get, as in (\ref{eq:limp}),
\begin{equation} \label{eq:limp2}
\gamma^\E  \leqslant c+2K \log(\epsilon(a)) \ll  \log(a)
\end{equation}
\\ 
Furthermore, we  call $T^{R}_{K, x}$ the restriction of $T^{R}$ to the box $[-Kx,(K+1)x]$, analogously as before.\\
 Bounding the eigenfunction correlator of $H^a$ with the Green function of a negative operator obtained by shifting $H^a$,  a  double resolvent expansion analogue to  the one performed in the proof of Theorem \ref{thm:2} and the Combes-Thomas bound yield
\begin{equation*}
\begin{split}
Q(0,x)\geqslant\,& \delta \kk G_{\delta}[H^a-(2+R)\cdot \one](0,x) \\
\geqslant \,& \delta \kk G_{\delta}[T^R- (2+R) \cdot \one](0,x)-C_3 \e^{-c_3(2K+1)\sqrt{\delta}|x| }
\end{split}
\end{equation*} 
on the event 
\begin{equation*}
\Omega_{x, \kk K; \kk a} := \{V_x^{\epsilon(a)}\equiv 0 \ \ \forall  \, x \in [-Kx,(K+1)x]\}, \ \quad  \PP\{\Omega_{x, \kk K; \kk a} \}=[\epsilon(a)]^{(2K+1)x}. 
\end{equation*}
Note that the fact that we are subtracting $(2+R)\cdot \one$ makes sure that $\delta$ is always above the spectrum of $H^{a}$. \\
Eventually, the only thing left to us to show is that
\begin{equation*}
G_{\delta}[T^{R}-(2+R) \cdot \one](0,x) \geqslant \widetilde{C}\e^{-\tilde{c} \delta |x|}~.
\end{equation*}
By writing down the Neumann series for $G_{\delta}[T^{R}-(2+R) \cdot \one] $ 
we get the following inequalities:
\begin{equation*}
\begin{split}
G_{\delta}[T^{R}-(2+R) \cdot \one]=\, & \dfrac{1}{\one- T_0(2+R+\delta-V^R)^{-1}} \cdot \dfrac{1}{2+R+\delta-V^R}  \\
= \, & \Big(\sum_{n=0}^\infty T_0^n(2+R+\delta-V^R)^{-n} \Big)\cdot \dfrac{1}{2+R+\delta-V^R} \\
\geqslant \, &  \dfrac{1}{2+\delta}\sum_{n=0}^\infty T_0^n(2+\delta)^{-n} \\
= \, & G_{\delta}[T_0- 2 \cdot \one] 
\end{split}
\end{equation*}
due to the positivity of $ T_0-V^R+(2+R) \cdot \one$. In particular
\begin{equation*}
G_{\delta}[T^R-(2+R) \cdot \one](0,x) \geqslant  G_{\delta}(T_0- 2 \cdot \one)] (0,x) \geqslant  \widetilde{C}\e^{-\tilde{c} \sqrt{\delta}\kk |x|}~.
\end{equation*}
Setting, again, $K$ large enough so that $2K+1 =\lceil 100 \kk \tilde{c}/c_3\rceil $, and setting $\delta =(\epsilon/\tilde{c})^2$, we finally conclude that 
\begin{equation*}
Q(0,x) \geqslant \delta G_{\delta}[T^R-(2+R) \cdot \one](0,x) \geqslant  \e^{-\epsilon |x|}.
\end{equation*}
This, combined with (\ref{eq:limp2}) and the Avron-Craig-Simon bound for general bounded potentials proven in Paragraph  \ref{sec:CS} implies the thesis.
\end{proof}

\end{document}